\documentclass[aps,prc,twocolumn,showpacs,nofootinbib,floatfix,superscriptaddress]{revtex4-1}
\usepackage{graphicx}
\usepackage{epsfig}
\usepackage{amsfonts}
\usepackage{amsmath}
\usepackage{hyperref}
\usepackage{float}
\usepackage{verbatim}
\usepackage{natbib}

\begin{document}

\title{Investigation of Heat Conductivity in Relativistic
Systems using a Partonic Cascade}

\author{M.\ Greif}
\email{greif@th.physik.uni-frankfurt.de}
\affiliation{Institut f\"ur Theoretische Physik,
Johann Wolfgang Goethe-Universit\"at,
Max-von-Laue-Str.\ 1, D-60438 Frankfurt am Main, Germany}

\author{F.\ Reining}
\affiliation{Institut f\"ur Theoretische Physik,
Johann Wolfgang Goethe-Universit\"at,
Max-von-Laue-Str.\ 1, D-60438 Frankfurt am Main, Germany}

\author{I.\ Bouras}
\affiliation{Institut f\"ur Theoretische Physik,
Johann Wolfgang Goethe-Universit\"at,
Max-von-Laue-Str.\ 1, D-60438 Frankfurt am Main, Germany}

\author{G.\ S.\ Denicol}
\affiliation{Department of Physics, McGill University, 3600 University Street, Montreal, Quebec H3A 2T8, Canada}

\author{Z.\ Xu}
\affiliation{Department of Physics, Tsinghua University, Beijing 100084, China}

\author{C.\ Greiner}
\affiliation{Institut f\"ur Theoretische Physik,
Johann Wolfgang Goethe-Universit\"at,
Max-von-Laue-Str.\ 1, D-60438 Frankfurt am Main, Germany}

%%%%%%%%%%%%%%%%%%%%%%%%%%%%%%%%%%%%%%%%%%%%%%%%%%%%%%%%%%%%%
%                          ABSTRACT
%%%%%%%%%%%%%%%%%%%%%%%%%%%%%%%%%%%%%%%%%%%%%%%%%%%%%%%%%%%%%

\begin{abstract}
Motivated by the classical picture of heat flow we construct a stationary
temperature gradient in a relativistic microscopic transport model. Employing the
relativistic Navier-Stokes ansatz we extract the heat
conductivity $\kappa$ for a massless Boltzmann gas using only binary collisions
with isotropic cross sections. We compare the numerical results to analytical
expressions from different theories and discuss the final results. The directly
extracted value for the heat conductivity can be referred to
as a literature reference within the numerical uncertainties.
\end{abstract}

\date{\today}
\maketitle

%%%%%%%%%%%%%%%%%%%%%%%%%%%%%%%%%%%%%%%%%%%%%%%%%%%%%%%%%%%%%
%                          INTRODUCTION
%%%%%%%%%%%%%%%%%%%%%%%%%%%%%%%%%%%%%%%%%%%%%%%%%%%%%%%%%%%%%

\section{Introduction}

Ultrarelativistic heavy ion collisions (HIC) at the Relativistic Heavy Ion
Collider (RHIC) and the Large Hadron Collider (LHC) create a hot and dense
system of strongly interacting nuclear matter that only existed in nature a
few microseconds after the Big Bang \cite{Rischke:2003mt,Tannenbaum:2006ch,Kolb:2003dz}.
At such energies, quarks
and gluons are deconfined and a new state of matter is formed, the
Quark-Gluon Plasma (QGP). The main experimental discovery made at RHIC and
LHC is that this novel state of matter behaves as a strongly coupled plasma,
with the smallest viscosity to entropy density ratio, $\eta /s$, ever
observed \cite{Arsene:2004fa,Adams:2005dq,Adcox:2004mh,Back:2004je}.
Currently, relativistic dissipative fluid dynamics is the main theory
employed to describe the space-time evolution of the QGP formed in HIC.

The inclusion of dissipative effects in the description of the QGP started
only a few years ago. To date the majority of studies have focused on investigating 
the effects of the shear viscosity in the time
evolution of the QGP and in extracting its magnitude from HIC measurements 
\cite{Heinz:2005bw,Romatschke:2007mq,Luzum:2008cw,Luzum:2012wu,Song:2010aq,Song:2010mg,Song:2011qa,Song:2011hk,Schenke:2010rr,Schenke:2011tv, Schenke:2011qd,Niemi:2011ix,Niemi:2012ry, Roy:2011xt}.
Nevertheless, there are other sources of dissipation that might play a
role in the fluid-dynamical description of HIC, such as bulk viscous
pressure and heat flow. While bulk viscous pressure effects have already been
subject to some amount of investigation \cite{Denicol:2009am,Denicol:2009pe,Denicol:2010tr,Monnai:2009ad,Monnai:2009dh,Schaefer:2012fc,Dusling:2011fd,Bozek:2009dw,Roy:2011pk,Roy:2012np, Song:2009rh},
the effects of heat flow have been largely ignored.

The main reason for neglecting heat flow is that most fluid-dynamical
calculations in the field attempt to describe the QGP only at midrapidity
and very high energies, where baryon number and its corresponding chemical
potential are approximately zero. However, when trying to describe the QGP
at forward rapidities or at smaller collision energies, such as the ones
probed in the RHIC low energy scan, baryon number can no longer be ignored
and heat conduction or baryon number diffusion might play a more decisive
role. On the other hand, before implementing heat flow in heavy ion
collision simulations, it is useful to study it with more detail in simpler
systems and check how well we are able to describe it in such cases. Such
types of studies were already started in Ref.\cite{Denicol:2012vq}.

In order to describe heat flow, one must know at least the heat conductivity
coefficient. Currently, this is not known with the desired precision even
for relativistic dilute gases. As a matter of fact, there are several
expressions in the literature for this transport coefficient, e.g. from
Israel-Stewart theory \cite{Israel}, resummed transient relativistic fluid
dynamics \cite{Denicol:2012cn}, or Chapman-Enskog theory \cite{Groot}. All
of these methods give slightly different results and it is useful to know which one
agrees best with the underlying microscopic theory in the dilute gas limit.
Recently, this task has been accomplished for the case of the shear
viscosity coefficient \cite{Wesp,Reining2}.

In this work, we investigate the heat flow of a stationary relativistic
dilute gas. Our main purpose is to obtain a precise expression for the heat
conduction coefficient in the kinetic regime. In Ref.~\cite{Reining2}, this
was done by imposing a stationary velocity gradient as a boundary condition,
and then waiting long enough for the system to achieve the Navier-Stokes
limit. Once the Navier-Stokes limit was obtained, the shear viscosity coefficient
was accurately extracted as the proportionality coefficient relating the
shear-stress tensor to the shear tensor. Here, we apply the method used in
Ref.~\cite{Reining2}, referred to as stationary gradients method, to extract
the heat conductivity of a dilute gas described by the relativistic
Boltzmann equation. We solve the relativistic Boltzmann equation numerically
using the transport model BAMPS (Boltzmann Approach for Multi-parton Scatterings)
described in Refs. \cite{Xu}. For the
sake of simplicity, we shall restrict our calculations to a classical gas of
massless particles, considering only elastic binary collisions with a
constant isotropic cross-section.

The paper is organized as follows: In Sec.~\ref{sec:basic_definitions} we
introduce the basic definitions of relativistic hydrodynamics. In the
following Sec.~\ref{sec:stationary_gradients} we give an overview of
stationary gradients. The numerical transport model BAMPS we use in this
work is introduced in Sec.~\ref{sec:BAMPS}. Next in Sec.~\ref%
{sec:anaGradients} we derive an analytical expression for the shape of the
gradients. In Sec.~\ref{sec:heatConductivity} we show the method to extract
the heat conductivity using BAMPS. The results obtained from the numerical
calculations are shown in Sec.~\ref{sec:results}, where we also discuss the
comparison to the analytical values. Finally, we give our conclusions in
Sec.~\ref{sec:conclusion}

Our units are $\hbar =c=k=1$; the space-time metric is given by $g^{\mu \nu
}=\text{diag}(1,-1,-1,-1)$.

%%%%%%%%%%%%%%%%%%%%%%%%%%%%%%%%%%%%%%%%%%%%%%%%%%%%%%%%%%%%%
%                          BASIC DEFINITIONS
%%%%%%%%%%%%%%%%%%%%%%%%%%%%%%%%%%%%%%%%%%%%%%%%%%%%%%%%%%%%%

\section{Basic Definitions}

\label{sec:basic_definitions}

In relativistic kinetic theory a dilute gas of particles is characterized by
the invariant particle distribution function $f(x,p)$. The macroscopic
quantities are obtained from the moments of this distribution function. The
first two moments of $f(x,p)$ correspond to currents of conserved
quantities: the first moment of $f(x,p)$ leads to the particle four-flow
(here we consider only binary collisions and, therefore, particle number is
conserved), 
\begin{equation}
N^{\mu }=\int \frac{gd^{3}p}{(2\pi )^{3}E}p^{\mu }f(p,x)\,,
\end{equation}%
while the second moment corresponds to the energy-momentum tensor, 
\begin{equation}
T^{\mu \nu }=\int \frac{gd^{3}p}{(2\pi )^{3}E}p^{\mu }p^{\nu }f(p,x)\,.
\end{equation}%
Above $g$ is the degeneracy factor, and $p^{\mu }=(E,\vec{p})$ is the
four-momentum of the particle. Since we consider the massless limit, $%
E=\left\vert \vec{p}\right\vert $.

The particle four-flow and energy-momentum tensor can be decomposed with
respect to an arbitrary normalized time-like four-vector, $u^{\mu }=\gamma
(1,\vec{v})\,$, where $u^{\mu }u_{\mu }=1$. The most general decomposition
\cite{Groot} reads 
\begin{align}
N^{\mu }& =nu^{\mu }+V^{\mu }, \\
T^{\mu \nu }& =\epsilon u^{\mu }u^{\nu }-P\Delta ^{\mu \nu }+W^{\mu }u^{\nu
}+W^{\nu }u^{\mu }+\pi ^{\mu \nu },
\end{align}%
where $\Delta ^{\mu \nu }=g^{\mu \nu }-u^{\mu }u^{\nu }$ is a projection
operator onto the 3-space orthogonal to $u^{\mu }$, $V^{\mu }$ is the
particle diffusion four-current, $W^{\mu }$ is the energy diffusion
four-current, and $\pi ^{\mu \nu }$ is the shear-stress tensor. Since we are
considering massless particles, the bulk viscous pressure is zero and was
omitted from the decomposition above. Furthermore, we introduced the local
rest frame (LRF) particle number density 
\begin{equation}
n\equiv N^{\mu }u_{\mu },
\end{equation}%
the LRF energy density 
\begin{equation}
\epsilon \equiv u_{\mu }T^{\mu \nu }u_{\nu }\,,
\end{equation}%
and the isotropic pressure
\begin{equation}
P=-\Delta _{\mu \nu }T^{\mu \nu }/3 \, .
\end{equation}%

Notice that the particle diffusion four-current can be written in terms of $%
N^{\mu }$ as 
\begin{equation}
V^{\mu }=\Delta _{\nu }^{\mu }N^{\nu },
\end{equation}%
while the energy diffusion 4-current can be expressed in terms of $T^{\mu
\nu }$ as, 
\begin{equation}
W^{\mu }=\Delta ^{\mu \alpha }T_{\alpha \beta }u^{\beta }.  \label{eq:Wmu}
\end{equation}%
The heat flow $q^{\mu }$ is defined as the difference between energy
diffusion and enthalpy diffusion. For the system we are considering, it
reads 
\begin{equation}
q^{\mu }=W^{\mu }-h\,V^{\mu }\;,
\end{equation}%
where $h=(\epsilon +P)/n$ is the enthalpy per particle.

Without loss of generality, we use Eckart's definition of the four-velocity 
\cite{Eckart:1940te}, 
\begin{equation}
u^{\mu }\equiv \frac{N^{\mu }}{\sqrt{N^{\mu }N_{\mu }}}.  \label{eq:Eckart}
\end{equation}%
By construction, the Eckart definition of the velocity field makes the
particle diffusion current vanish, $V^{\mu }=0$, and, consequently, the heat
flow in the Eckart frame is exactly given by the energy diffusion, i.e, 
\begin{equation}
q^{\mu }=W^{\mu }.  \label{eq:qmu_eckart}
\end{equation}
We remark that the definition of the velocity field will not affect the
value of the heat conduction coefficient.

For a gas of classical particles, the thermodynamic pressure and particle
number density are connected to the temperature, $T$, as follows, 
\begin{equation}
p=nT.  \label{eq:idealGasEquation}
\end{equation}%
Since we consider a massless Boltzmann gas, the thermodynamic pressure
$p$ is equal to the the isotropic pressure $P$. The fugacity,
$\lambda \equiv \exp \left( \mu /T\right) $, is given by 
\begin{equation}
\lambda =\frac{n}{n_{\mathrm{eq}}},  \label{eq:fugacity}
\end{equation}%
where $n_{\mathrm{eq}}=gT^{3}/\pi ^{2}$ is the equilibrium particle number
density with vanishing chemical potential.

%%%%%%%%%%%%%%%%%%%%%%%%%%%%%%%%%%%%%%%%%%%%%%%%%%%%%%%%%%%%%
%                        STATIONARY TEMPERATURE GRADIENTS
%%%%%%%%%%%%%%%%%%%%%%%%%%%%%%%%%%%%%%%%%%%%%%%%%%%%%%%%%%%%%

\section{Stationary Temperature gradients}

\label{sec:stationary_gradients}

In Navier-Stokes theory \cite{Groot} heat flow is proportional to the gradient of the
thermal potential, $\alpha =\mu /T$, 
\begin{align}
q^{\mu }& =-\kappa \frac{nT^{2}}{\epsilon +p}\bigtriangledown ^{\mu }\alpha \nonumber
\\
& =\kappa \left( \bigtriangledown ^{\mu }T-\frac{T}{\epsilon +p}%
\bigtriangledown ^{\mu }p\right) \,,  \label{heatflowformel}
\end{align}%
where $\kappa $ is defined as the heat conduction coefficient, $\partial
_{\mu }$ is the ordinary four-derivative, $D=u^{\mu }\partial _{\mu }$ is
the comoving derivative, and $\bigtriangledown ^{\mu }\equiv \Delta ^{\mu
\nu }\partial _{\nu }=\partial ^{\mu }-u^{\mu }D$ is the space-like
four-gradient. For the sake of simplicity, we assume the system to be
homogeneous in the $y$ and $z$ plane (here referred to as the transverse
direction) and resolve only the dynamics in the $x$--direction. We also wait
long enough so that the system achieves a stationary solution, i.e, $%
\partial _{t}=0$. In such stationary limit and with the spatial symmetries
described above, the four-derivative simplifies to $\partial _{\mu
}=(0,\partial _{x},0,0)$.

In the stationary limit, conservation of momentum dictates that the pressure
gradient must vanish. Furthermore, for small velocities, $u^{x}\ll 1$, the
operator $\bigtriangledown ^{\mu }$ reduces to 
\begin{equation}
\bigtriangledown ^{\mu }=\left( 
\begin{array}{c}
u_{0}u_{x}\partial _{x} \\ 
-\partial _{x} \\ 
0 \\ 
0%
\end{array}%
\right) .
\end{equation}%
Then the $x$-component of the heat flow can be cast in the following simple
form 
\begin{equation}
q^{x}=\kappa \left( \bigtriangledown ^{x}T\right) = - \kappa \partial _{x}T(x).
\label{eq:heatFlowGradientExpression}
\end{equation}%
In this simplified scenario, the heat conduction coefficient can be
extracted as the proportionality coefficient between the heat flow and the
gradient of temperature.

%%%%%%%%%%%%%%%%%%%%%%%%%%%%%%%%%%%%%%%%%%%%%%%%%%%%%%%%%%%%%
%                          BAMPS
%%%%%%%%%%%%%%%%%%%%%%%%%%%%%%%%%%%%%%%%%%%%%%%%%%%%%%%%%%%%%

\section{The partonic cascade BAMPS}

\label{sec:BAMPS}

In this work, the relativistic Boltzmann equation is solved numerically
using the BAMPS simulation, developed
and previously employed in Refs.~\cite{Xu,Bouras:2009nn,Bouras,Reining2,Fochler,Uphoff:2011ad}.
This partonic cascade solves the Boltzmann equation, 
\begin{equation}
\label{eq:BTE}
p^{\mu }\partial _{\mu }f(x,p)=C\left[ f\right] ,
\end{equation}%
for on-shell particles using the stochastic interpretation of transition
rates. In this study we consider only binary collisions with constant isotropic
cross sections. In order to reduce statistical fluctuations in simulations and to
ensure an accurate solution of the Boltzmann equation \eqref{eq:BTE}
a testparticle method \cite{Xu} is introduced: The particle number
is artificially increased by multiplying it by the number of
testparticles per real particle, $N_{\rm test}$. The physical results
are not affected by this step.

\begin{figure}[h]
\centering
\includegraphics[width=0.5\textwidth]{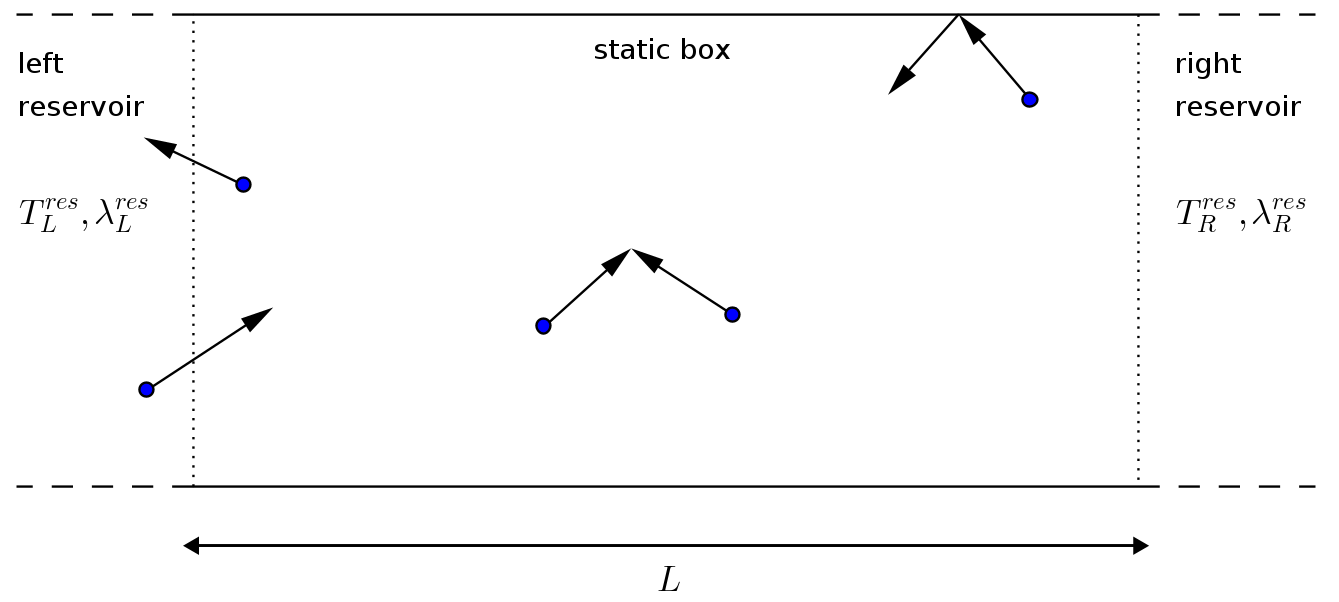}
\caption{A sketch of the setup in BAMPS. In the left and right sides ($x$%
--direction) of the static box we introduce thermal reservoirs with fixed
chemical potential and temperature.}
\label{fig:sketchBAMPS}
\end{figure}

The simulation of the relativistic Boltzmann equation is performed in a
static box. The transverse plane ($y$--$z$ plane) of the system is assumed
to be homogeneous. This is maintained by realizing the collisions of particles against
the boundaries of the static box as elastic collisions off a
wall. In the $x$--direction the boundaries of the box are fixed to have a
constant gradient in temperature. In practice, this is implemented in BAMPS
by removing the particles colliding with the wall in the $x$-direction.
Independent of the absorption, the reservoirs emit particles  with fixed
temperature and chemical potential. The reservoirs in the left and right boundaries of
the box in the $x$--direction, as sketched in Fig~\ref{fig:sketchBAMPS},
are defined to have fixed temperatures, $%
T_{L}^{\mathrm{res}}$ and $T_{R}^{\mathrm{res}}$, and fixed fugacities, $%
\lambda _{L}^{\mathrm{res}}$ and $\lambda _{R}^{\mathrm{res}}$,
respectively. Their values in our calculations are chosen to
$T_{L}^{\mathrm{res}}=0.5\,\mathrm{GeV}$, $T_{R}^{\mathrm{res}}=0.3\,\mathrm{GeV}$.
In order to maintain no
pressure gradient, we set the fugacities to $\lambda _{L}^{\mathrm{res}}=1.0$,
and $\lambda _{R}^{\mathrm{res}}=7.72$. The chosen values in the
reservoirs imply that, $p_{L}^{\mathrm{res}} = p_{R}^{\mathrm{res}} = 13.25 \,\mathrm{GeV/fm^3}$.
The collective velocity in both reservoirs is set to zero.

The set of boundary conditions will lead to a
stationary temperature gradient in the $x$--direction. Then the heat
conduction coefficient can be extracted using Eq.~%
\eqref{eq:heatFlowGradientExpression}. The required time for the system to
reach its stationary state depends on value of the cross section, the
temperature and chemical potential of the thermal reservoirs, and the
initial state of the system.

Similar to the work in \cite{Reining2}, the time the profile needs to
reach the stationary profile is proportional to the inverse mean free path.
This implies that the more diffuse the system is, the faster it reaches
the final stationary profile and vice versa. In order to save computational time, we
always initialize the system as close as possible to its stationary state.

The first step in computing the heat flow is extracting the components of
the particle four-flow $N^{\mu }$ and energy-momentum tensor $T^{\mu \nu }$.
Space in BAMPS is discretized in small volume elements, $V_{c}$.
The distribution function $f(x,p)$ of this volume element is
reconstructed from the momenta distribution of the particles inside it. In
this scheme, the $N^{\mu }$ and $T^{\mu \nu }$ are computed via the discrete
summation over all particles within the specific volume element and divided
by the testparticle number: 
\begin{align}
N^{\mu }(t,x)& =\frac{1}{V_{c} N_{\rm test}}\sum\limits_{i=1}^{N_{c}}\frac{p_{i}^{\mu }}{%
p_{i}^{0}},  \label{eq:nmu_BAMPS} \\
T^{\mu \nu }(t,x)& =\frac{1}{V_{c} N_{\rm test}}\sum\limits_{i=1}^{N_{c}}\frac{p_{i}^{\mu
}p_{i}^{\nu }}{p_{i}^{0}},  \label{eq:tmunu_BAMPS}
\end{align}%
where $N_{c}$ is the total number of particles inside the corresponding
volume, $t$ is the time, and $x$ is the space coordinate (defined to be in
the center of the volume element). Since we shall be considering only
stationary solutions, the $t$ dependence of the current will be
ommited in the following sections. We also mention here, that we
average \eqref{eq:nmu_BAMPS} and \eqref{eq:tmunu_BAMPS} over many
events in order to reduce statistical fluctuations.

Using Eqs.~\eqref{eq:nmu_BAMPS} and \eqref{eq:tmunu_BAMPS} we can
determine all necessary macroscopic quantities. Then, using Eqs.~%
\eqref{eq:nmu_BAMPS} and \eqref{eq:Eckart}, we compute the four-velocity in
the Eckart frame and extract the particle number density, $n$, and the
projection operator $\Delta ^{\mu \nu }$. Finally, we compute the heat flow
current using Eqs.~\eqref{eq:Wmu} and \eqref{eq:qmu_eckart}.

%%%%%%%%%%%%%%%%%%%%%%%%%%%%%%%%%%%%%%%%%%%%%%%%%%%%%%%%%%%%%
%                         ANALYTIC EXPRESSION GRADIENTS
%%%%%%%%%%%%%%%%%%%%%%%%%%%%%%%%%%%%%%%%%%%%%%%%%%%%%%%%%%%%%

\section{Analytical expression for the gradients}

\label{sec:anaGradients}
In this work we aim to extract the gradient of a macroscopic quantity,
such as the temperature $T$ or the particle density $n$.
The gradients are usually computed using
finite-difference methods which demands a large amount of statistics in
order to obtain a sufficiently smooth profile. Therefore, if one is able to
obtain the general form of the solution that such gradients must satisfy in
the stationary regime, it would save a huge amount of computational runtime.

As already shown in \cite{Reining2},
quantities that are conserved in collisions show a linear behavior
between the thermal reservoirs.
For the boundary conditions implemented in this work, this is realized for  
the particle density.

%%Instead of:
%In the case of a temperature gradient
%and only binary collisions as realized in this work, this is true
%for the particle density. 

For the type of boundary condition considered
in this work, the particle
number density satisfies the following solution in the stationary regime 
\begin{equation}
n(x)=\frac{n_{R}^{\mathrm{res}}-n_{L}^{\mathrm{res}}}{L+2\lambda _{\mathrm{%
mfp}}}\,x+\frac{n_{R}^{\mathrm{res}}+n_{L}^{\mathrm{res}}}{2},
\label{eq:gradientDensityAnalytic}
\end{equation}%
where $n_{L}^{\mathrm{res}}$ and $n_{R}^{\mathrm{res}}$ are the particle
number density in the left and right reservoirs, respectively, and $L$ is
the size of the static box in the $x$ direction. We remark that this
solution might be modified when inelastic collisions are included.

\begin{figure}[t!]
\includegraphics[width=0.5\textwidth]{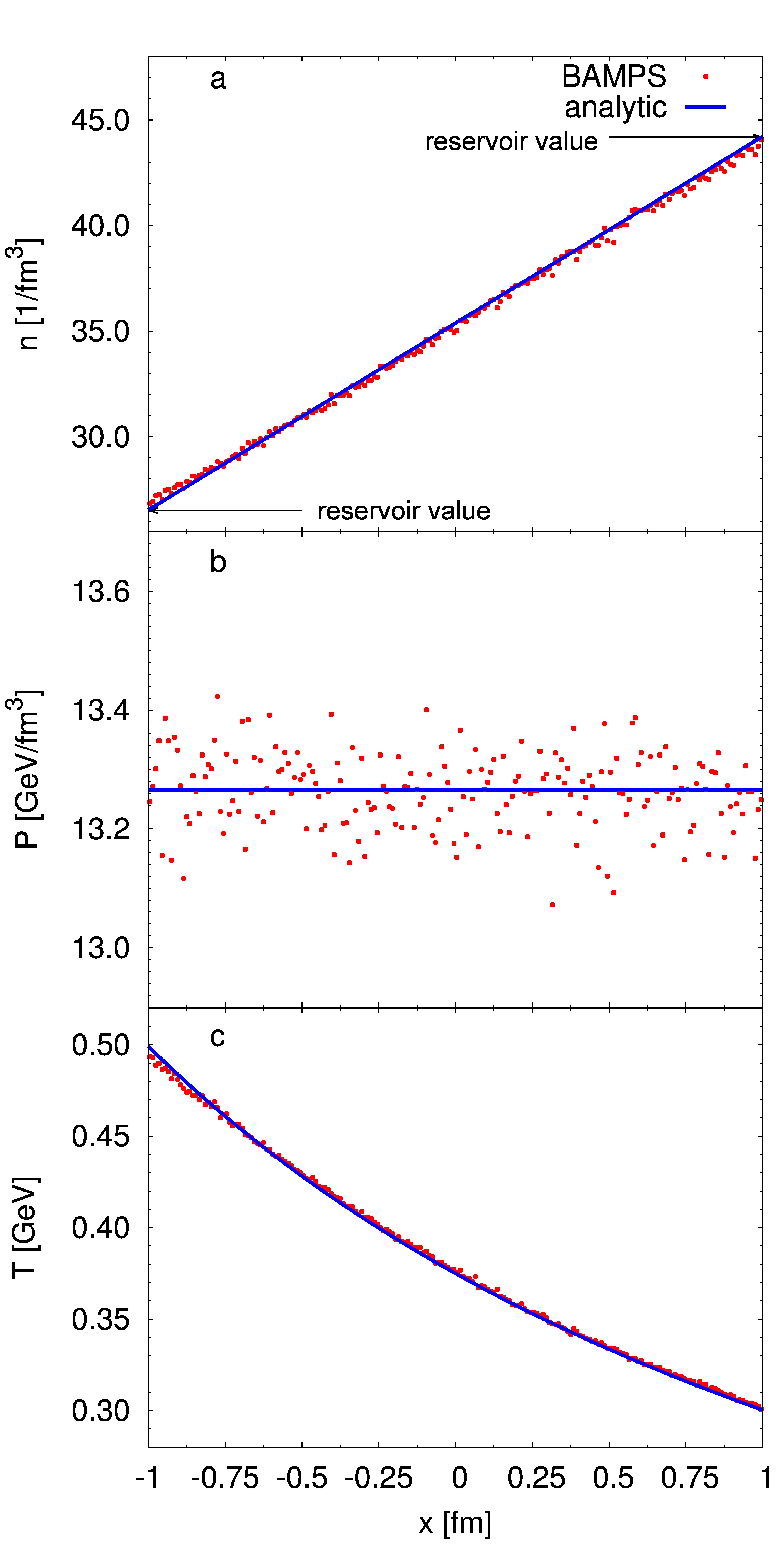}
\caption{The numerical solutions extracted from BAMPS for particle density $%
n $, pressure $P$ and temperature $T$. The cross section is set to $\protect%
\sigma _{22} =10\,\rm mb$. The solutions are shown at very late times, $t=50\,%
\mathrm{fm/c}$, where the profiles are almost static. The reservoir values
are chosen on the left (right) to $T_{L}^{res}=0.5\,\mathrm{GeV}$ and $%
\protect\lambda _{L}^{res}=1.0$ ($T_{R}^{res}=0.3\,\mathrm{GeV}$ and $%
\protect\lambda _{R}^{res}=7.72$). We compare each profile with the
analytical results derived from Eq.~\eqref{eq:gradientDensityAnalytic}.
The results are averaged over 500 events.}
\label{fig:initialProfiles}
\end{figure}
In general, the above solution does not hold near the boundaries (in the $x$%
--direction) of the static box, where the particle number density is
discontinuous (see Ref. \cite{Reining2} for details). If the mean free-path, 
$\lambda _{\mathrm{mfp}}$, of the particles is large compared to the size of
the box, $L$, such discontinuity will lead to deviations from the above
solution. Here we tackle this problem by making sure that $L\gg \lambda _{%
\mathrm{mfp}}$. This will reduce such finite-size effects to the minimum
amount possible and guarantees that the solution in Eq.~%
\eqref{eq:gradientDensityAnalytic} provides a very good description of the
particle number density in most parts of the system. Using Eq.~%
\eqref{eq:gradientDensityAnalytic} and the fact that the thermodynamic
pressure is constant inside the box, it is straightforward to obtain the
temperature profile,
\begin{equation}
T\left( x\right) =p\left( \frac{n_{R}^{\mathrm{res}}-n_{L}^{\mathrm{res}}}{%
L+2\lambda _{\mathrm{mfp}}}\,x+\frac{n_{R}^{\mathrm{res}}+n_{L}^{\mathrm{res}%
}}{2}\right) ^{-1}.  \label{eq:gradientTAnalytic}
\end{equation}
In Fig.~\ref{fig:initialProfiles}, we show the particle number density (a),
thermodynamic pressure (b), and temperature (c) profiles computed with
BAMPS (dots) at $t=50$ \textrm{fm} and
$\sigma _{22} = 10 \, \rm mb$, with the
boundary conditions specified in the previous section. The blue curves
correspond to the analytical solution described above. It is clear that Eq.~%
\eqref{eq:gradientDensityAnalytic} does in fact reproduce the solutions
obtained numerically with BAMPS and that finite-size effects are negligible.
Also, we confirm that the thermodynamic pressure is constant, as it should be.

\section{Extraction of the heat conductivity}

\label{sec:heatConductivity}

Since the gradients of the particle number density and temperature can be
computed analytically from Eqs.~\eqref{eq:gradientDensityAnalytic}
and \eqref{eq:gradientTAnalytic}, and the pressure $p$ is by construction constant,
the extraction of the heat conductivity
from BAMPS simulations becomes straightforward. Using $n(x) \equiv ax + b$
for the particle density and the relation \eqref{eq:idealGasEquation} we
can simplify Eq.~\eqref{eq:heatFlowGradientExpression} to
\begin{equation}
q^x= \kappa p \frac{a}{(ax+b)^2} \, .
\end{equation}
Finally, the heat conductivity can be calculated using 
\begin{equation}
\label{eq:heatFlowCoefficientBAMPS}
 \kappa= q^x  \cdot \, \frac{(ax+b)^2}{a p} \, ,
\end{equation}
where the heat flow $q^{x}$ is directly extracted from
BAMPS and the constants $a$ and $b$ as well as the constant
pressure $p$ are analytically known.

\begin{figure}[h]
\includegraphics[width=0.5\textwidth]{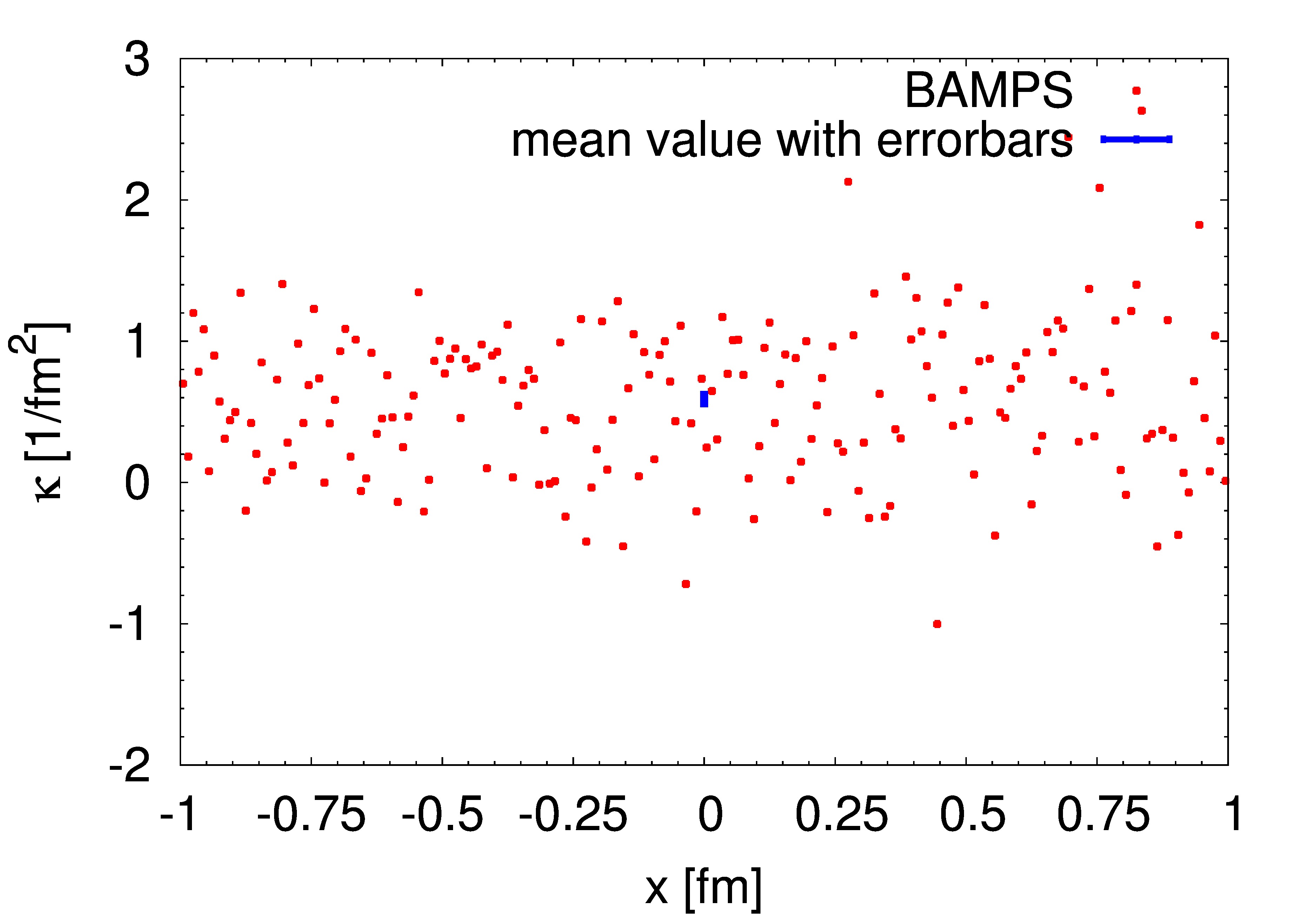}
\caption{Heat conductivity coefficient extracted from BAMPS shown at every
point in space. The mean value with error bar is also displayed.
The results are shown at $t=5\,\mathrm{fm/c}$ for $\protect%
\sigma _{22}=43\,\mathrm{mb}$. 500 events were averaged.}
\label{fig:kappa_space}
\end{figure}

\begin{figure}[h]
\includegraphics[width=0.5\textwidth]{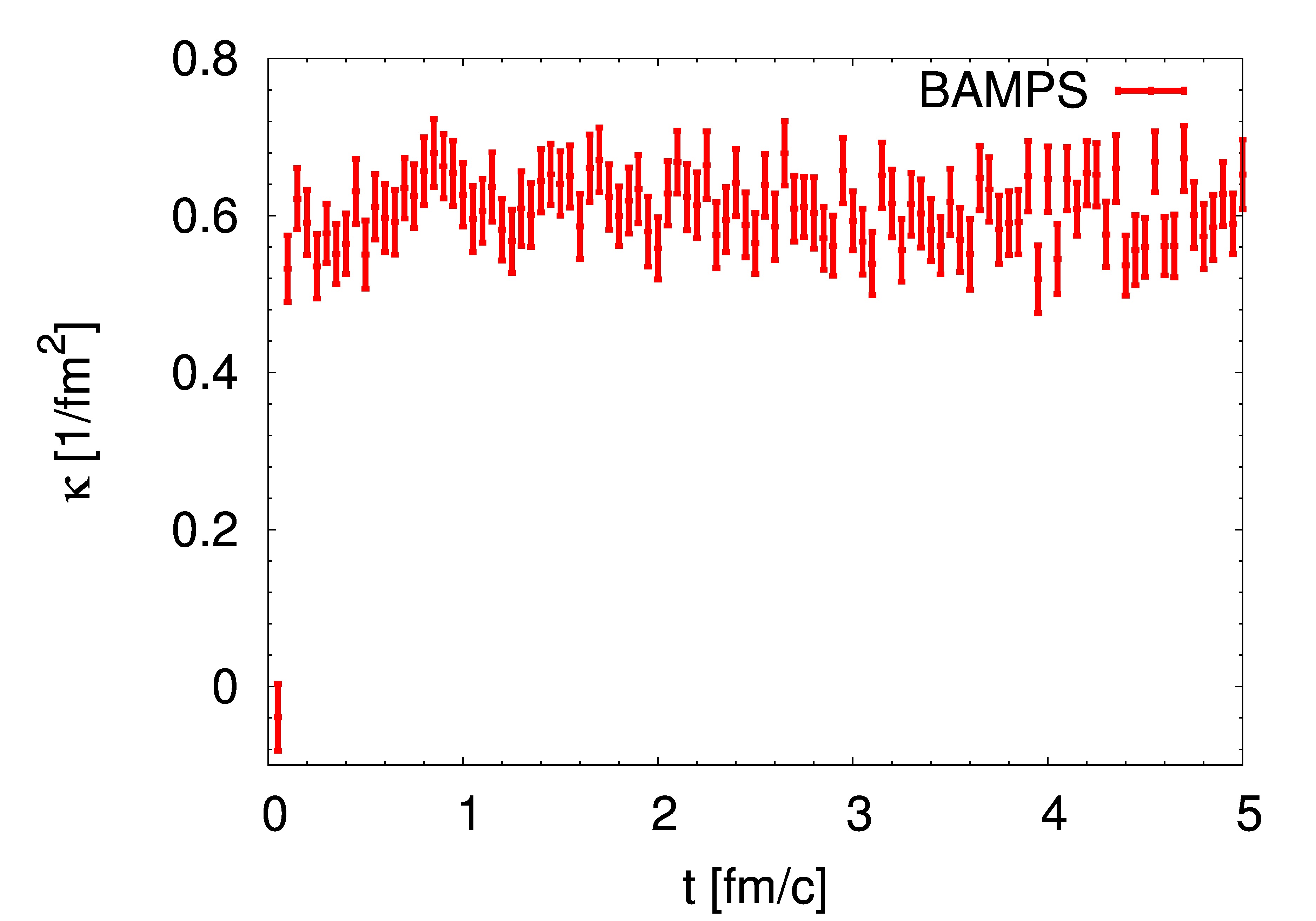}
\caption{Heat conductivity coefficient extracted from BAMPS shown at
different times averaged over the all positions in space. The results are
are shown for $\protect\sigma _{22} =43\,\mathrm{mb}$. 500 events were averaged.}
\label{fig:kappa_time}
\end{figure}

Figure \ref{fig:kappa_space} shows the heat conductivity as function of $x$,
at $t=5$ \textrm{fm}. We use for the cross section $\sigma _{22} =43\,\rm mb$.
It is clear that $\kappa $ is constant over space,
which confirms that we reached the asymptotic solution.
In Fig.~\ref{fig:kappa_time} we show the heat conductivity
as function of time $t$. The system relaxes locally on a very short
time scale to its stationary solution, and the heat conductivity is
afterwards basically constant over time. We remark, that the relaxation
time is proportional to the mean free path, $\lambda_{\rm mfp}$
(see Ref.~\cite{Denicol:2012cn}).

In order to extract a precise value for heat conductivity $\kappa $,
we take an average over the values obtained in the whole system and at all
time steps that had reached the asymptotic state. 

%%%%%%%%%%%%%%%%%%%%%%%%%%%%%%%%%%%%%%%%%%%%%%%%%%%%%%%%%%%%%
%                          RESULTS
%%%%%%%%%%%%%%%%%%%%%%%%%%%%%%%%%%%%%%%%%%%%%%%%%%%%%%%%%%%%%

\section{Results}

\label{sec:results}

In kinetic theory, there are several different methods to compute the
transport coefficients appearing in the fluid-dynamical equations of motion,
e.g. Chapman-Enskog theory and the many variations of the method of moments.
The problem is that each method predicts different expressions for the
transport coefficients. Even though these differences are not so large for
the shear viscosity coefficient, which has been discussed in details in
Refs.~\cite{Wesp,Reining2}, they can be significantly large for heat
conductivity, as will be discussed in the following. The results from BAMPS
can help to clarify which method reflects more reliably the underlying
microscopic theory.

The method of moments, initially developed by H. Grad for nonrelativistic
systems \cite{Grad} and further extended to describe relativistic systems by
several authors, is one of the most commonly employed methods in heavy ion
collisions. Traditionally, this method is employed in the relativistic
regime together with the so-called 14-moment approximation, originally
proposed by Israel and Stewart \cite{Israel}. In this scheme, the momentum
distribution function, $f(x,p)$, is expanded in momentum space around its
local equilibrium value in terms of a series of Lorentz tensors formed of
particle four-momentum $k^{\mu }$, i.e., $1,\,k^{\mu },\,k^{\mu }k^{\nu
},\,\ldots $. The 14-moment approximation consists in truncating the
expansion at second order in momentum, i.e., keeping only the tensors $%
1,\,k^{\mu }$, and $k^{\mu }k^{\nu }$ in the expansion, leaving 14 unknown
expansion coefficients. The coefficients of the truncated expansion are then 
uniquely matched to the 14 components of $N^{\mu }$ and $T^{\mu \nu }$.

Israel and Stewart obtained the equations of motion of fluid dynamics and,
consequently, the microscopic expressions of the transport coefficients
appearing in such equations, by substituting the momentum distribution
function truncated according to the 14-moment approximation into the second
moment of the Boltzmann equation, 
\begin{equation*}
\int \frac{d^{3}k}{k^{0}}k^{\mu }k^{\nu }k^{\lambda }\partial _{\mu }f=\int 
\frac{d^{3}k}{k^{0}}k^{\nu }k^{\lambda }C\left[ f\right] .
\end{equation*}%
By projecting this equation with $\Delta _{\nu }^{\alpha }u_{\lambda }$,
Israel and Stewart obtained an equation of motion for the heat flow and a
microscopic formula for $\kappa $ \cite{Israel, Denicol:2012es}. For a
massless and classical gas of hard spheres, one obtains the following
expression for the heat conductivity \cite{Denicol:2012es, Groot},%
\begin{equation}
\kappa =\frac{2}{\sigma _{22}}.  \label{lauciello-equation}
\end{equation}

Note that Israel-Stewart's 14-moment approximation leads to ambiguous
results since it can be substituted in any moment of the Boltzmann equation 
\cite{Denicol:2012es}. By changing the moment in which the 14-moment
approximation is replaced, the equations of motion remain with the same
general form, but the transport coefficients become different. For example,
using the choice of moment applied in Ref.~\cite{Denicol:2010xn}, one
obtains a quantitatively different result for $\kappa $,%
\begin{equation}
\kappa =\frac{3}{\sigma _{22}}.
\label{three-over-sigma-equation}
\end{equation}%
Naturally, other choices of moment will lead to even more different results,
but we shall not list them all here.

Recently, the derivation of fluid dynamics from the method of moments was
extended in order to remove the ambiguities of the 14-moment approximation\ 
\cite{Denicol:2012cn}. The main difference between the 14-moment
approximation and the theory derived in Ref.\ \cite{Denicol:2012cn},
namely Resummed Transient Relativistic Fluid Dynamics (RTRFD), is that
the latter does not truncate the moment expansion of the momentum
distribution function. Instead, dynamical equations for all its moments are
considered and solved by separating the slowest microscopic time scale from
the faster ones. The resulting fluid-dynamical equations are truncated
according to a systematic power-counting scheme using the inverse Reynolds
and the Knudsen numbers. The values of the transport coefficients of fluid
dynamics are obtained by re-summing the contributions from all moments of
the momentum distribution function, similar to what happens in
Chapman-Enskog theory \cite{CE}. In practice, the transport coefficients are
given by a summation of terms, each term originating from a specific moment
of $f(x,p)$. Once the transport coefficients have converged, the summation
can be truncated and the moments that are irrelevant can be dropped out. For
the case of a massless and classical gas of hard spheres, the first three
moments are enough to obtain a convergent value for $\kappa $, and one
obtains \cite{Denicol:2012cn} 
\begin{equation}
\kappa =\frac{2.5536}{\sigma _{22}}.  \label{denicol-equation}
\end{equation}%
Including the next moment would only lead to $\sim 1\%$ corrections to this
expression.

Finally, heat conductivity can also be computed using Chapman-Enskog theory,
as was done in Ref.~\cite{Groot} for a gas of classical and massless hard
spheres,
\begin{equation}
\label{chapmanEnskogg}
\kappa =\frac{2.44}{\sigma _{22}}.
\end{equation}%
Note that the convergence of the above value was not investigated in Ref. 
\cite{Groot} and, consequently, it is not possible to infer how precise this
result is.

We found excellent agreement to the result of Ref.~\cite{Denicol:2012cn},
i.e., Eq.\eqref{denicol-equation}.
\begin{figure}[h]
\includegraphics[width=0.5\textwidth]{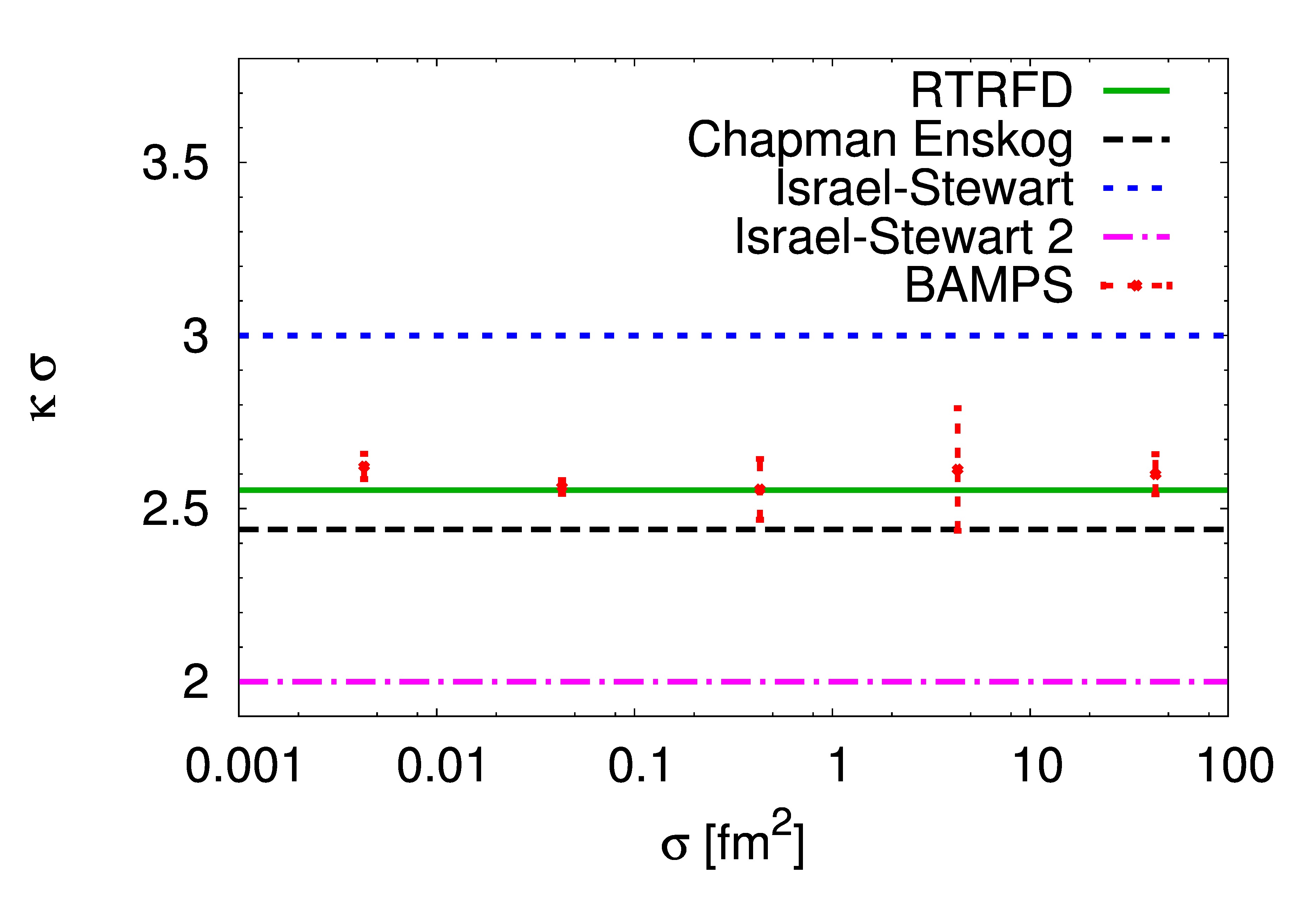}
\caption{The dimensionless quantity $\kappa \sigma$ derived from
different fluid dynamical theories (lines) compared to the results extracted
directly from BAMPS (dots).}
\label{fig:finalpicture}
\end{figure}
This can be seen in Fig.~\ref%
{fig:finalpicture}, where results obtained using BAMPS are compared to Eqs.~%
\eqref{lauciello-equation}, \eqref{three-over-sigma-equation}, \eqref{denicol-equation} and  \eqref{chapmanEnskogg}. The original
Israel-Stewart theory \eqref{lauciello-equation} yields values roughly $\sim
22\%$ too low, while the 14-moment approximation with the choice of moment
described in Ref.~\cite{Denicol:2010xn} leads to a result that is $\sim 17\%$
too high. The transport coefficient computed via Chapman-Enskog theory is
close to the value obtained from microscopic theory, being $\sim 4\%$ below
it. However, it is not clear whether this can be considered as a failure of
the Chapman-Enskog theory since the precision of that result was not clearly
stated and it may be possible that it can be improved to match what was
extracted from the Boltzmann equation. Finally, the heat conductivity
coefficient computed in Ref.~\cite{Denicol:2012cn} was basically the same as
the one computed by BAMPS, within the errors of both the theoretical
calculation (originating from the truncation of the moment expansion in the
expression for $\kappa $) and the numerical one (originating from the finite
statistics of the calculation).
\begin{table}[t]
\centering
\begin{tabular}{|c|c|}
\hline
$\sigma_{22}$ & $\kappa~[1/fm^2]$ \\ \hline
$0.043~mb$ & $609.695698\pm 8.4513$ \\ \hline
$0.43~mb$ & $59.588731\pm 0.4694$ \\ \hline
$4.3~mb$ & $5.943535\pm 0.2041$ \\ \hline
$43~mb$ & $0.607672\pm 0.0411$ \\ \hline
$430~mb$ & $0.060449\pm 0.0013$ \\ \hline
\end{tabular}
\caption{Numerical results for the heat conductivity coefficient $\protect%
\kappa $ for various elastic cross sections over several orders of
magnitude.\label{Tabelle_1}}
\end{table}
Nevertheless, all theoretical calculations predicted a $1/\sigma _{22}$
dependence of the heat conductivity coefficient on the cross section; a fact
also confirmed by BAMPS, see Tab.~\ref{Tabelle_1} and Fig.~\ref{fig:finalpicture}.

\section{Conclusion and outlook}

\label{sec:conclusion}

In this work we extracted the heat conductivity coefficient for a dilute gas
of massless and classical particles described by the relativistic Boltzmann
equation. For this purpose we employed the microscopic transport
model BAMPS in a static system using only binary collisions and a constant
isotropic cross section. For this setup, we established a stationary temperature
gradient using thermal reservoirs. In order to simplify the
calculations we derived
an analytical expression of the expected profile to obtain the gradients of
the temperature, meanwhile the heat flow was extracted directly from BAMPS
by the general decomposition of the particle four-flow and energy momentum
tensor. Using the relativistic Navier-Stokes theory, which is valid for
a stationary system and small gradients, we extracted the heat conductivity
to a very high precision. We then compared this result with
several theoretical predictions for this transport coefficient, each
originating from a different derivation of fluid dynamics from the
underlying microscopic theory.

The numerical simulations of the Boltzmann equation, realized in this paper
with BAMPS, were able to distinguish between all these different theoretical
results and clearly point out which one was in better agreement with the
underlying microscopic theory. While Israel-Stewart theory, i.e., the
14-moment approximation, performed rather poorly in the description of heat
flow, new extensions of the method of moments, i.e. RTRFD, were able to provide
an improved description of this transport coefficient, showing a very good
agreement with the coefficient extracted from BAMPS. The heat conductivity
computed using Chapman-Enskog theory was not able to precisely describe the
simulation from BAMPS, although it was not as far off as the predictions
using the 14-moment approximation. Nevertheless, this disagreement might be
originating from a poor implementation of the Chapman-Enskog theory, which
was not properly checked for the convergence of the transport coefficient.

The extracted value for the heat conductivity
\begin{equation}
\kappa_{\rm BAMPS} =\frac{2.59 \pm 0.07}{\sigma _{22}}.
\end{equation}
can be referred to as a literature value, within the numerical uncertainties,
for the simple case of binary collisions with an isotropic cross section.
It remains as a future task to extend this work to include inelastic
scatterings, where particle production and annihilation have to be taken into
account.

\section{Acknowledgments}

The authors are grateful to the Center for the Scientific Computing (CSC) at
Frankfurt for the computing resources. FR and IB are grateful to ``Helmhotz
Graduate School for Heavy Ion research''. FR and ZX acknowledges supported by BMBF. 
GSD is supported by the Natural Sciences and Engineering Research Council of Canada.
ZX is supported by the NSFC under grant No. 11275103.
This work was supported by the Helmholtz International Center for FAIR within
the framework of the LOEWE program launched by the State of Hesse.

\bibliographystyle{ieeetr}
\bibliography{Literatur}

\end{document}